\documentclass[conference]{IEEEtran}
\usepackage[square,sort&compress,comma,numbers]{natbib}
\usepackage{color}
\usepackage{graphicx}
\usepackage{amsfonts}
\usepackage{epstopdf}
\usepackage{latexsym}
\usepackage{amssymb}
\usepackage{amsmath}
\usepackage{multirow}
\usepackage{enumerate}
\usepackage{threeparttable}
\usepackage{cancel}

\usepackage{tikz,xcolor,hyperref}
\definecolor{lime}{HTML}{A6CE39}
\DeclareRobustCommand{\orcidicon}{%
    \begin{tikzpicture}
    \draw[lime, fill=lime] (0,0) 
    circle [radius=0.16] 
    node[white] {{\fontfamily{qag}\selectfont \tiny ID}};    \draw[white, fill=white] (-0.0625,0.095) 
    circle [radius=0.007];    \end{tikzpicture}
    \hspace{-2mm}}
\foreach \x in {A, ..., Z}{%
    \expandafter\xdef\csname orcid\x\endcsname{\noexpand\href{https://orcid.org/\csname orcidauthor\x\endcsname}{\noexpand\orcidicon}}
    }
\hypersetup{hidelinks,
	colorlinks=true,
	allcolors=black,
	pdfstartview=Fit,
	breaklinks=true}

\RequirePackage{amsmath,amsthm,amsfonts,amssymb,bm,mathrsfs}
\IEEEoverridecommandlockouts

\begin{document}

\title{Multi-Channel Multi-Domain based Knowledge Distillation Algorithm for Sleep Staging with Single-Channel EEG}
\author{\IEEEauthorblockN
        {Chao Zhang\orcidA{},
        Yiqiao Liao,
        Siqi Han,
        Milin Zhang\orcidB{},~\IEEEmembership{Senior Member,~IEEE},\\
        Zhihua Wang\orcidC{},~\IEEEmembership{Fellow,~IEEE},
        Xiang Xie\orcidD{}
        }

\thanks{Chao Zhang and Yiqiao Liao contributed equally to this work.}
\thanks{Chao Zhang and Milin Zhang are with the Department of Electronic Engineering, Institute for Precision Medicine, and Beijing National Research Center for Information Science and Technology, Tsinghua University, Beijing, China, 100084. Corresponding author e-mail: zhangmilin@tsinghua.edu.cn.}
\thanks{Yiqiao Liao, Zhihua Wang and Xiang Xie are with the School of Integrated Circuits, Tsinghua University, Beijing, China, 100084.}
\thanks{Siqi Han is with the School of Modern Post (School of Automation), Beijing University of Posts and Telecommunications, Beijing, China, 100876.}
\thanks{This work is supported in part by the National Key Research and Development Program of China (No.2018YFB220200*), in part by the Natural Science Foundation of China through grant 92164202. in part by the Beijing Innovation Center for Future Chip, in part by the Beijing National Research Center for Information Science and Technology.}
\thanks{Digital Object Identifier}
}

\maketitle

\begin{abstract}
This paper proposed a Multi-Channel Multi-Domain (MCMD) based knowledge distillation algorithm for sleep staging using single-channel EEG. Both knowledge from different domains and different channels are learnt in the proposed algorithm, simultaneously. A multi-channel pre-training and single-channel fine-tuning scheme is used in the proposed work. The knowledge from different channels in the source domain is transferred to the single-channel model in the target domain. A pre-trained teacher-student model scheme is used to distill knowledge from the multi-channel teacher model to the single-channel student model combining with output transfer and intermediate feature transfer in the target domain. The proposed algorithm achieves a state-of-the-art single-channel sleep staging accuracy of 86.5\%, with only 0.6\% deterioration from the state-of-the-art multi-channel model. There is an improvement of 2\% compared to the baseline model. The experimental results show that knowledge from multiple domains (different datasets) and multiple channels (e.g. EMG, EOG) could be transferred to single-channel sleep staging.
\end{abstract}

\begin{IEEEkeywords}
Sleep staging, Transfer learning, Knowledge distillation, Single-channel EEG, Brain-computer interface
\end{IEEEkeywords}

\IEEEpeerreviewmaketitle

\section{Introduction}
\IEEEPARstart{S}{leep} staging is an essential technique for sleep-related disease diagnosis and treatment. According to the sleep staging definition by the American Academy of Sleep Medicine (AASM), there are 5 sleep stages: Wake, Non-Rapid Eye Movement 1 (N1), N2, N3, and Rapid Eye Movement (REM).
The golden standard for sleep staging is manual labelling on Polysomnography (PSG) signals by doctors. The PSG signals consist of Electroencephalography (EEG), Electrooculography (EOG), Electromyography (EMG), and Electrocardiogram (ECG). A considerable amount of literature \cite{phan2019seqsleepnet,chambon2018deep,phan2018joint,zhang2017new} has been published on automatic sleep staging using PSG signals with feasible accuracy.
Seqsleepnet \cite{phan2019seqsleepnet} achieved a state-of-the-art accuracy based on multi-channel signals for sleep staging using a sequence-to-sequence hierarchical recurrent neural network (RNN). However, the acquisition of PSG signals involves bulky equipment, which limits the potential scenarios that can be applied.

In recent years, various wearable sleep monitoring devices have been released using single-channel EEG signal for sleep status analysis. Different single-channel EEG signal based sleep staging algorithms have been reported in literature \cite{supratak2017deepsleepnet,phan2020towards,liao2020design,phan2018automatic,liao2020tri,phan2018dnn,chang2019ultra,cai2021graph,liao2021lightsleepnet,fiorillo2021deepsleepnet,hassan2017automated,hassan2017decision}. 
DeepSleepNet \cite{supratak2017deepsleepnet} combines the time-invariant features from Convolutional Neural Network (CNN) and temporal features from Bidirectional Long Short-Term Memory (Bi-LSTM) for single-channel sleep staging.
However, the insufficiency of the single-channel data for training is a big issue in improving the accuracy. The widely applied deep learning models are easy to be over-fitting while the dataset is too small. As the amount of dataset is fixed, the practical method to increase the available training data is to either introduce information from other domains into the training set or take the typical ignored information into account.

In order to introduce information from other domains into the training set, one choice is to get more knowledge from a bigger dataset. \cite{luo2018eeg} proposed a Conditional Wasserstein Generative Adversarial Network (GAN) framework to generate EEG data for data augmentation, but the GAN failed to create knowledge that does not exist in the dataset. Recently, pre-trained representation model such as Bidirectional Encoder Representations from Transformers (BERT) \cite{devlin2018bert} achieved state-of-the-art performance for natural language processing tasks. Following this trend, \cite{phan2020towards} proposed a transfer learning approach to transfer knowledge from a large dataset to a small cohort. The model was pre-trained in the source domain, and then was fine-tuned in the target domain. It achieved the state-of-the-art accuracy in single-channel sleep staging.
However, the cross-channel knowledge transfer is ignored.

\begin{figure*}[!tb]
\centering
\includegraphics[width=0.7\textwidth]{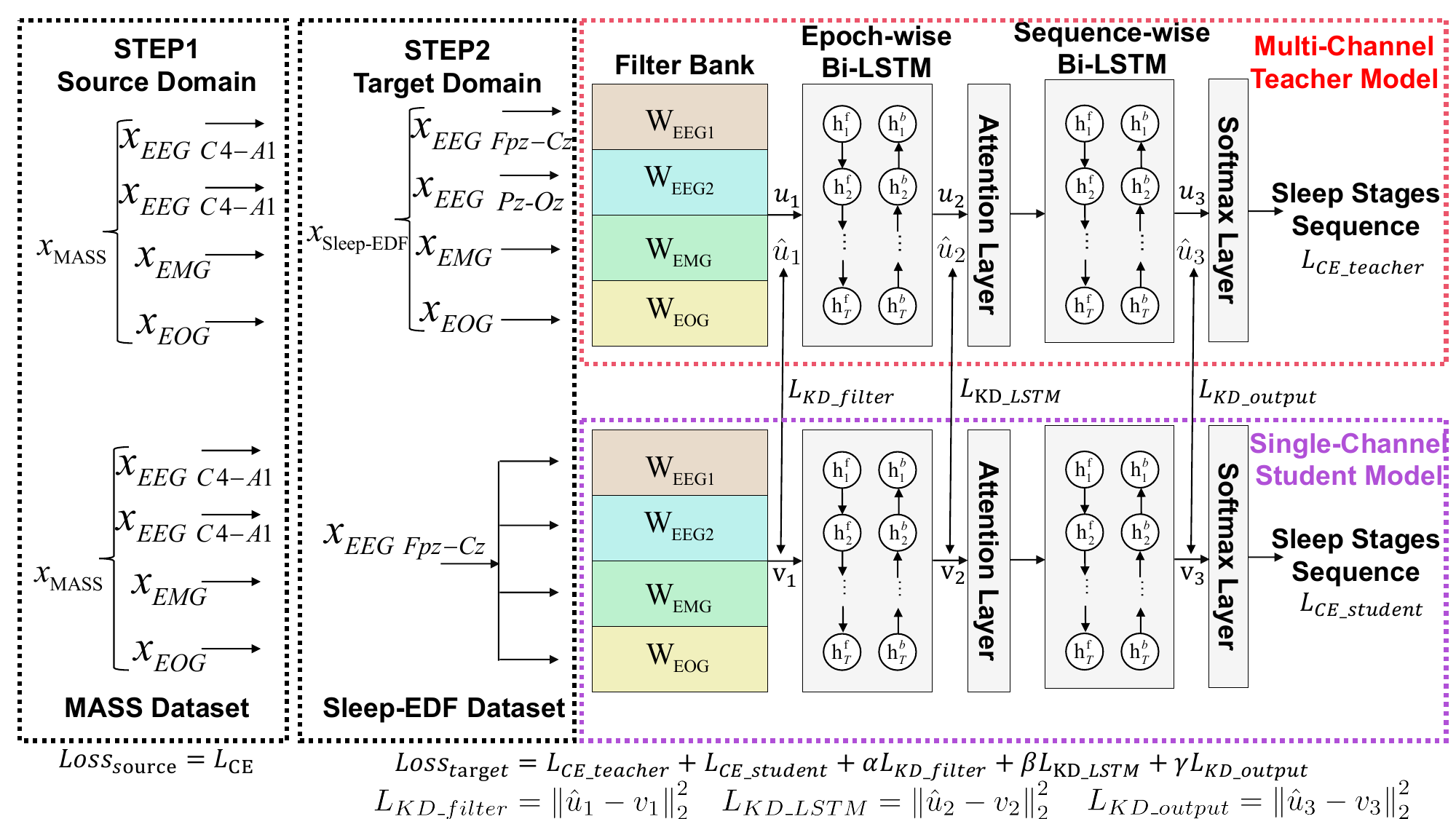}
\caption{The overall architecture of the proposed MCMD algorithm. It consists of two steps: 1) Source Domain Pre-training. 2) Target Domain Knowledge Distillation. In the step 1, four-channel model $M_0$ is trained using the three-channel MASS dataset. In the step 2, $M_0$ is used to initialize the teacher model $M_T$ and the student model $M_S$. $M_T$ takes multi-channel signals $x_{sleep-edf}$ as its input, while $M_S$ only takes $x_{EEG Fpz\_Cz}$ as its input. The teacher and student models are trained simultaneously in the Sleep-EDF dataset while knowledge distillation is utilized between the filterbank, LSTM, and output of these two models.}
\label{fig:architecture}
\end{figure*}

Another promising solution is to utilize knowledge from the ignored channels. The concept of knowledge distillation \cite{hinton2015distilling} was proposed for model compression. The teacher model, $M_T$, usually features higher accuracy but with higher complexity, while the student model, $M_S$, features lower accuracy with a light-weight architecture. $M_S$ learns from $M_T$ with satisfactory accuracy and appropriate complexity.
Our previous work \cite{liao2020design} proposed a competition and cooperation based knowledge distillation model, where $M_T$ is a multi-channel model, and $M_S$ is a single-channel model. It enhances the performance of single-channel EEG with knowledge transfer between channels. Yet, the knowledge in other domains is ignored. 

This paper proposed a Multi-Channel Multi-Domain (MCMD) based knowledge distillation algorithm for single-channel EEG based sleep staging. The domains consist of the MASS dataset \cite{o2014montreal} as source domain and the Sleep-EDF dataset \cite{kemp2000analysis,goldberger2000physiobank} as target domain. The channels consist of EEG, EMG and EOG channels. The proposed algorithm combines knowledge transfer in four different scenarios: Same-Domain Same-Channel (SDSC), Same-Domain Cross-Channel (SDCC), Cross-Domain Same-Channel (CDSC) and Cross-Domain Cross-Channel (CDCC). It consists of two steps: 1) Source domain pre-training. 2) Target domain knowledge distillation. In the first step, the pre-trained model from the source domain was used for initializing the teacher and the student models. Both CDSC and CDCC transfer was employed for those models. In the second step, knowledge distillation was applied from the multi-channel teacher model to the single-channel student model with the combination of output transfer, feature transfer, and filterbank transfer. SDSC and SDCC knowledge transfer was employed in this step.
The proposed algorithm achieves an accuracy of 86.5\%, which is higher than the previous single-channel sleep staging works reported on literature. There is only a 0.4\% deterioration from our multi-channel teacher model, as well as a 2\% improvement compared with the baseline Seqsleepnet \cite{phan2019seqsleepnet}. The experimental results show the effectiveness of the four knowledge transfer scenarios. Knowledge from different channels, or even different domains could be used in single-channel sleep staging.

The rest of this paper is organized as follows: Section II introduces the proposed MCMD knowledge distillation algorithm with the experimental results shown in Section III, while Section IV concludes our work.

\section{MCMD based knowledge distillation}
\figurename \ref{fig:architecture} illustrates the architecture of the proposed MCMD knowledge distillation algorithm. The training of the proposed algorithm consists of two steps: 1) Source domain pre-training. 2) Target domain knowledge distillation. The proposed algorithm takes four-channel signals from the source domain (MASS dataset) and the target domain (Sleep-EDF dataset) for training. It employs a multi-channel teacher model and a single-channel student model for sleep staging in the target domain. 

The Seqsleepnet model is chosen for $M_T$ and $M_S$. The input of the model is time-frequency representations of 30s PSG epochs. A short-time Fourier transform is applied to transform the 30s PSG raw data into power spectra with a number of frequency bins, $F$, of 129, a number of time indices, $T$, of 29, and a number of channels, $C$, of 4.
The structure of the model consists of filterbank layers, epoch-wise Bi-LSTM with a input length of the frame number in one epoch, attention layers, sequence-wise Bi-LSTM  with a input length of the epoch number in one sequence, and Softmax layers.

\subsection{Source Domain Pre-Training}
The source domain is MASS dataset consisting of EEG C4-A1, EMG, and EOG.
The target domain is Sleep-EDF consisting of EEG Fpz-Cz, EEG Pz-Oz, EMG, and EOG. The EEG channel C4-A1 is duplicated for the corresponding of the EEG Pz-Oz channel in the target domain,  
which eliminates the channel number mismatch and increases the number of cross domain knowledge transfer paths.
The Seqsleepnet model $M_0$ is trained with sequence Cross-Entropy loss function $H(y,p)$, which is defined as:
\begin{equation}
H(y,p) =  - \frac{1}{{{N_b}}}\frac{1}{{{L}}}\sum\limits_{k = 1}^{{N_b}}\sum\limits_{j = 1}^{{L}} {\sum\limits_{i = 1}^5 {{y_{ji}^{k}}} } \log ({p_{ji}^{k}})
\end{equation}
where $y_{ji}^{k}$ denotes the $i^{th}$ sleep stage in the one-hot ground truth label of the $j^{th}$ sample in the $k^{th}$ sequence of the corresponding batch. $L$ is the length of a sequence, which is set as 30. $N_b$ is the size of one batch. $p$ is the prediction outputs after Softmax activation which can be calculated as:
\begin{equation}\label{eq:loss_P}
\begin{aligned}
p= {\rm{softmax}}(u({x_{C4-A1},x_{C4-A1},x_{EMG},x_{EOG}}))\\
\end{aligned}
\end{equation}
where $u(*)$ denotes the last hidden layer output from $M_0$.
The best Seqsleepnet model $M_0$ for the four channels with parameters $W_0$ can be found as:
\begin{equation}\label{eq:weight}
\begin{aligned}
W_0 = \mathop {argmin}\limits_{{W_0}} (H(y,p(x_{C4-A1},x_{C4-A1},x_{EMG},x_{EOG})) )
\end{aligned}
\end{equation}

$M_T$ and $M_S$ will be both initialized from $M_0$. Thus, multi-channel knowledge is transferred from the source domain to the target domain. $M_T$ would be used for all the signals from Sleep-EDF dataset including EEG Fpz-Cz, EEG Pz-Oz, EMG and EOG. However, $M_S$ is a single channel model only using EEG Fpz-Cz. Although the number of the  input channels is different for $M_T$ and $M_S$, they share the same network structure and initialization model $M_0$. $M_T$ and $M_S$ are then fine-tuned in the target domain for knowledge transfer. As $M_S$ requires input from four channels, the single-channel input EEG Fpz-Cz is duplicated for three times. 

There are two types of transfer in the first step: 1) Cross-Domain Same-Channel (CDSC) and 2) Cross-Domain Cross-Channel (CDCC). The knowledge from the EEG C4-A1 is transferred to similar channels such as EEG Fpz-Cz and EEG Pz-Oz from other domain by applying the CDSC transfer. In particular, with the CDCC transfer, the EMG and EOG channels from other domain would also be transferred to the EEG channels.

\subsection{Target Domain Knowledge Distillation}

The multi-channel knowledge in the target domain is even more crucial for the single-channel sleep staging as there is no domain shift. There is a knowledge distillation between $M_T$ and $M_S$ after the pre-training. The $M_T$ model can learn from the multi-channel signals and teach the $M_S$ model.
The loss function for this knowledge distillation is defined as:
\begin{equation}\label{eq:loss-target}
\begin{aligned}
Loss_{target} = L_{CE\_teacher}+L_{CE\_student }+\alpha L_{KD\_filter}
\\+\beta L_{KD\_LSTM}+\gamma L_{KD\_output}
\end{aligned}
\end{equation}
where $L_{CE\_teacher}$ and $L_{CE\_student }$ are sequence cross-entropy loss functions for $M_T$ and $M_S$, respectively. $L_{KD\_filter}$, $L_{KD\_LSTM}$ and $L_{KD\_output}$ are knowledge distillation losses of the filterbank, the epoch-wise LSTM, and the hidden layer output between the teacher and the student, respectively. The hyper-parameters $\alpha$, $\beta$ and $\gamma$ are all set as 1500.
The loss of the knowledge distillation is defined as:
\begin{equation}\label{eq:loss-filter}
\begin{aligned}
L_{KD\_filter}=\left \| \hat u_1-v_1\right \|_{2}^{2}
\end{aligned}
\end{equation}
\begin{equation}\label{eq:loss-lstm}
\begin{aligned}
L_{KD\_LSTM}=\left \| \hat u_2-v_2\right \|_{2}^{2}
\end{aligned}
\end{equation}
\begin{equation}\label{eq:loss-output}
\begin{aligned}
L_{KD\_output}=\left \| \hat u_3-v_3\right \|_{2}^{2}
\end{aligned}
\end{equation}
where $\hat u$ and $v$ are output features after the filterbank, the epoch-wise LSTM and the final hidden layer for $M_T$ and $M_S$, respectively.

Simultaneous training is applied to the proposed teacher-student system. A more robust model can be expected since $M_S$ could learn from the dynamic paths for training instead of a static pre-trained $M_T$ with fixed parameters. However, the knowledge distillation loss increases the similarity between $M_T$ and $M_S$, which may result in a reduction in the accuracy of $M_T$ and $M_S$ simultaneously. A gradient block is applied to the knowledge distillation loss to stop $M_S$ decreasing the performance of $M_T$ during the simultaneous training.

\begin{figure}[htbp]
\centering
\includegraphics[width=0.5\textwidth]{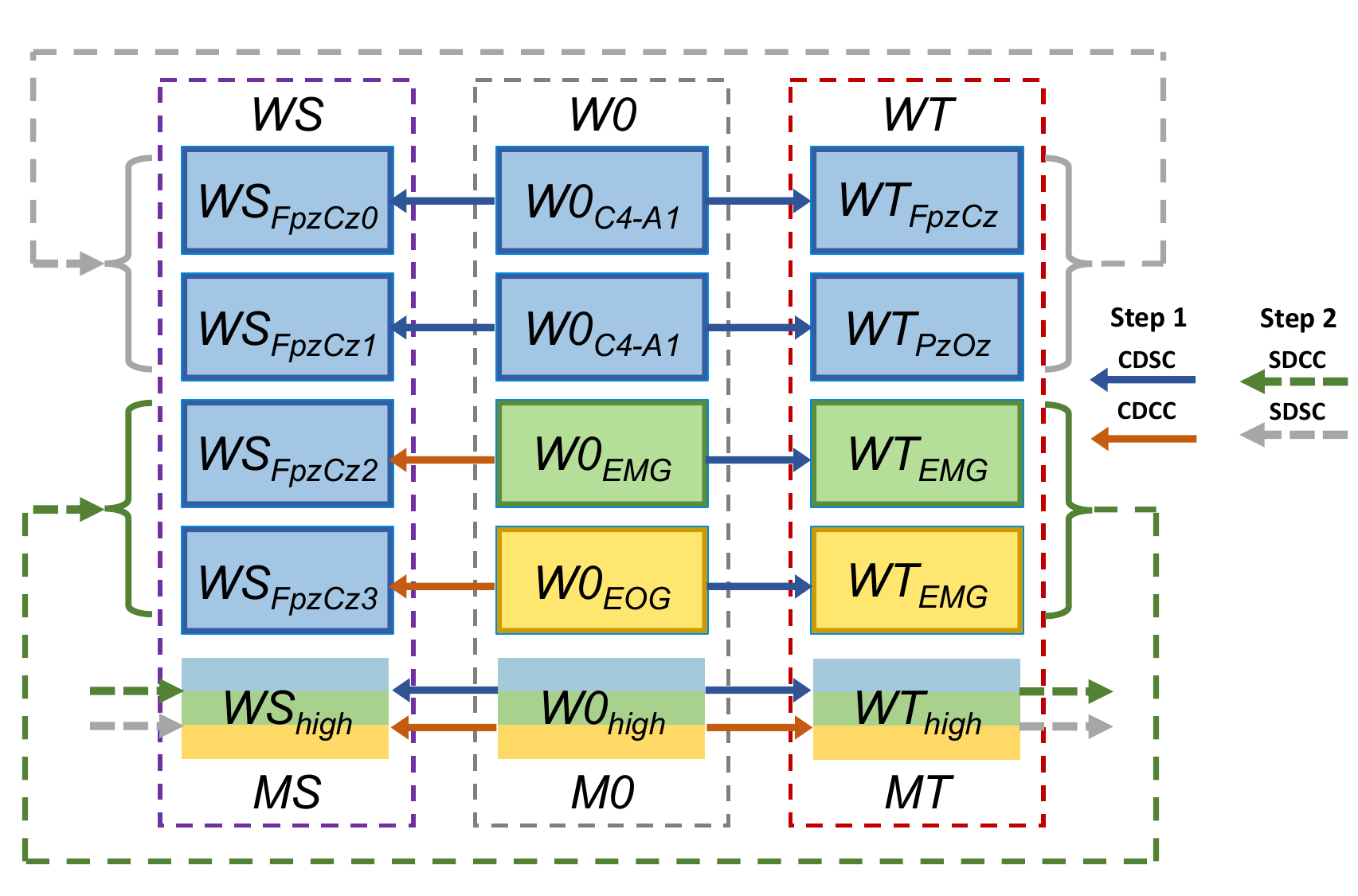}
\caption{Knowledge Transfer Paths in the proposed work. There are four types of knowledge transfer methods: CDSC, CDCC, SDCC, and SDSC.}
\label{fig:transfer}
\end{figure}

\figurename \ref{fig:transfer} illustrates the knowledge transfer paths in the proposed work. There are four types of knowledge transfer methods: CDSC, CDCC, SDCC, and SDSC. The model is divided into two parts: the low-level filterbank and the high-level network. The filterbank consists of fully connected layers. The network consists of a hierarchical Bi-LSTM, an attention layer and a fully connected layer. In the low-level filterbank, each weight is responsible for one channel independently. The weights in the high-level network deal with all the four channels simultaneously. In the first step, there are CDSC and CDCC knowledge transfer between $M_0$ and $M_T$, $M_S$ with the pre-training and fine-tuning scheme. The knowledge could directly be transferred between the models through the different weights. For CDSC, the knowledge is transferred between similar channels from different domains. For instance, the EEG C4-A1 of the MASS dataset is transferred to the EEG Fpz-Cz of the Sleep-EDF dataset. Between these two domains, there is a significant difference due to the usage of different acquisition devices. In addition, the locations of the EEG channels are also different. However, as they share a low level representation of EEG, a simple duplication of the weight in $M_0$ still provides a proper initialization for the training, which prevents the model from over-fitting with the increase of the amount of training data. For CDCC, the knowledge origins from different channels of different domains. For example, the EMG and EOG of the source domain are transferred to the EEG channels in the target domain. EMG and EOG are meaningful for sleep staging since the EMG signal is usually used in the classification among the REM and NREM stages, and the EOG helps determine when sleep occurs as well as whether the subject is in the REM sleep stage. The weights of the two channels, $W0_{EMG}$ and $W0_{EOG}$, contain information that is useful in the determination of those sleep stages. Thus the introduction of the EMG and EOG channels is helpful for the training of $M_S$ and $M_T$.

In the second step, there are SDCC and SDSC transfer between $M_T$ and $M_S$. The knowledge distillation loss forces the intermediate layers of these two networks to output the same results. The knowledge distillation loss also causes an indirect knowledge transfer between the weights. There is knowledge transfer from the EMG and EOG channels to the EEG channel in SDCC transfer, as well as from the two EEG channels to the single-channel EEG Fpz-Cz with SDSC transfer. Since $M_T$ is trained based on multi-channel signals, it converges to a smoother feature space. Even if the weights responsible for the single-channel EEG processing are taken out separately, it would be better than the models trained using the single-channel EEG. Therefore, in SDSC, besides EEG Pz-Oz, EEG Fpz-Cz in $M_T$ will also be transferred to the same channel in $M_S$. There are three different knowledge distillation losses monitored in this step. Firstly, $L_{KD\_output}$ pushes the output of the final hidden layer from the two models be similar to each other. The output is a posterior probability distribution for sleep staging. As not disturbed by the number of channels, it is the most effective knowledge distillation approach. Secondly, $L_{KD\_filter}$ compels these two models to output equal time-invariant features extracted by the filterbank, which is equivalent to the translation of different channel features.
Thirdly, $L_{KD\_LSTM}$ makes the epoch-wise LSTM of the outputs from $M_T$ and $M_S$ share the same temporary feature. It is a supplement for time-invariant feature translation. In the combination of the filterbank transfer, the LSTM feature transfer, and the output transfer, the knowledge distillation loss enforces the two models similar not only in output, but also in the features of the intermediate layers with different inputs and weights. In this case, $M_S$ learns the relationship between the different channels, although the input signal is only EEG Fpz-Cz, the intermediate features are similar to $M_T$ with a four-channel input. 

\section{Experimental Results}
The proposed work applies pre-training in the source domain using the MASS dataset. The pre-training was completed under a 20-fold cross validation protocol for 100 epochs. Then the model with the highest accuracy was retained. The knowledge distillation is applied in the target domain using the Sleep-EDF dataset. The Sleep-EDF dataset is used for testing by a leave-one-subject-out 20-fold cross-validation. The MASS dataset contains 200 participants with overnight EEG records and corresponding sleep stages. The dataset contains one scalp-EEG signals from C4-A1 channel, a submental chin EMG, and a horizontal EOG. All the three channels were used for the pre-training of the four-channel model $M_0$. The Sleep-EDF dataset contains 20 participants with two scalp-EEG signals from Fpz-Cz and Pz-Oz channels, an EMG, and an EOG. For the multi-channel teacher model, signals from all four channels are employed, while only the EEG Fpz-Cz channel is employed for the single-channel student models. The sleep stages consist of Wake, N1, N2, N3 and REM.

\begin{table}[htbp]
  \centering
  \caption{Comparison between Transfer Methods}
  \scriptsize
    \begin{tabular}{|l|l|l|l|}
    \hline
    \multicolumn{3}{|c|}{Transfer Methods} & \multirow{2}*{ACC} \\
    \cline{1-3}
    Output Transfer & Filterbank Transfer & LSTM Transfer & ~ \\
    \hline
    $\checkmark$ & $\checkmark$ &$\checkmark$&86.52\%\\
    \hline
    $\checkmark$ & $\times$ &$\times$&86.44\%\\
    \hline
    $\times$ & $\checkmark$ &$\times$&86.31\%\\
    \hline
    $\times$ & $\times$ &$\checkmark$&86.26\%\\
    \hline
    $\times$ & $\times$ &$\times$&85.77\%\\
    \hline
    \end{tabular}%
  \label{tb:transfer}%
\end{table}%

Table \ref{tb:transfer} compares the performance between different transfer methods from the proposed work. 
It is noted that the proposed output transfer features the best, since the posterior probability distribution from the final hidden layer contains more information concerning sleep staging without being contaminated by other channels. In addition, with the combination of these three transfer losses, a better result is achieved. The output of the filterbank contains more useful knowledge than the LSTM, and this might because that the filterbank transfer enables one-to-one channel transfer by extracting features for different channels independently using different weights. In contrast, the LSTM transfer employs all the four channels simultaneously. The results also indicate that any transfer method is better than nothing.

\begin{table}[!tp]
  \caption{Comparison With The State-of-The-Art}
  \label{tb:comparision}
  \scriptsize
\begin{tabular}{|l|l|l|l|l|l|l|}
\hline
\multirow{2}{*}{Methods} & \multirow{2}{*}{Dataset} & \multirow{2}{*}{Channels} & \multirow{2}{*}{\begin{tabular}[c]{@{}l@{}}Transfer\\ Methods\end{tabular}} & \multicolumn{3}{c|}{Overall Metrics} \\ \cline{5-7} & & & & ACC & MF1 & kappa \\ 
\hline
\textbf{This work} & \begin{tabular}[c]{@{}l@{}}Sleep-EDF\\(MASS)\end{tabular} & Single & \begin{tabular}[c]{@{}l@{}l@{}l@{}}CDSC\\CDCC\\SDSC\\SDCC\end{tabular} & \textbf{86.5} & 80.9 & 0.82 \\ 
\hline
TNSRE18 \cite{chambon2018deep} & Sleep-EDF & Single & None & 81.4 & 72.2 & - \\ 
\hline
TBE18 \cite{phan2018joint} & Sleep-EDF & Single & None & 81.9 & 73.8 & 0.74 \\ 
\hline
TNSRE17 \cite{supratak2017deepsleepnet} & Sleep-EDF & Single & None & 82.0 & 76.9 & 0.76 \\ \hline
EMBC18 \cite{phan2018automatic} & Sleep-EDF & Single & None & 82.5 & 72.0 & 0.76 \\ 
\hline
EMBC18 \cite{phan2018dnn} & Sleep-EDF & Single & None & 82.6 & 74.2 & 0.76 \\ 
\hline
ISCAS20 \cite{liao2020design} & Sleep-EDF & Single & \begin{tabular}[c]{@{}l@{}}SDSC\\SDCC\end{tabular} & 83.7 & 75.7 & 0.78 \\ 
\hline
TCASII21 \cite{liao2021lightsleepnet} & Sleep-EDF & Single & None & 83.8 & 75.3 & 0.78 \\ 
\hline
TBE20 \cite{phan2020towards} & \begin{tabular}[c]{@{}l@{}}Sleep-EDF\\ (MASS)\end{tabular} & Single & CDSC & 85.2 & 79.6 & 0.79 \\ 
\hline
TNSRE21 \cite{fiorillo2021deepsleepnet} & Sleep-EDF & Single & None & 86.1 & 79.2 & 0.81 \\ 
\hline
TNSRE19 \cite{phan2019seqsleepnet} & MASS & Three & None & 87.1 & 81.5 & 0.833 \\ 
\hline
\end{tabular}
\end{table}

\newcommand{\tabincell}[2]{\begin{tabular}{@{}#1@{}}#2\end{tabular}}
\begin{table}[htbp]
  \centering
  \caption{Ablation Experiments in MCMD Algorithm}
  \scriptsize
    \begin{tabular}{|l|l|l|l|l|}
    \hline
    \multicolumn{5}{|c|}{Student Model} \\
    \hline
    Transfer Method&Source Domain & Target Domain &Capacity&ACC \\
    \hline
    Baseline-1C&No & Single-Channel &1C&84.57\%\\
    \hline
    CDSC&Single-Channel & Single-Channel &1C&85.61\%\\
    \hline
    CDSC+CDCC&Multi-Channel & Single-Channel &4C&85.77\%\\
    \hline
    \tabincell{c}{CDSC+CDCC\\
    +SDSC+SDCC }&Multi-Channel & Multi-Channel &4C&86.52\%\\
    \hline
    \multicolumn{5}{|c|}{Teacher Model} \\
    \hline
    Transfer Method&Source Domain & Target Domain &Capacity&ACC \\
    \hline
    Baseline&No & Multi-Channel &4C&86.27\%\\
    \hline
    CDSC+CDCC&Multi-Channel & Multi-Channel &4C&86.91\%\\
    \hline
    \end{tabular}%
  \label{tb:ablation}%
\end{table}%

Table \ref{tb:comparision} compares the proposed work with previous works. The MCMD model achieves an accuracy (ACC) of 86.5\% and a Macro-F1 (MF1) of 80.9 in single-channel EEG sleep staging. With the combination of four types of transfer methods, our model achieves a comparable result with multi-channel model using only single-channel EEG. 
An improvement of 0.4\% is achieved while comparing to the state-of-the-art single channel sleep staging method in literature \cite{fiorillo2021deepsleepnet} based on subject independent 20-fold cross validation.
Compared with our previous work \cite{liao2020design}, there is a 2.8\% improvement of ACC and a 5.4 improvement of MF1 from the introduction of multi-domain knowledge, which also indicates that it solves the problem of unbalanced sample distribution.

Table \ref{tb:ablation} demonstrates the Ablation experiment results of the MCMD algorithm. Different knowledge transfer methods and model capacities are applied. The baseline Seqsleepnet model achieves an ACC of 84.57\%.
According to \cite{phan2020towards}, there is a 1.04\% improvement compared with the baseline by using single-channel pre-training from the source domain. The proposed multi-channel pre-training features better ACC by applying a combination of CDSC and CDCC. As only a small part of the knowledge transfer is taken across different channels and domains, the improvement is not high. With a teacher-student knowledge distillation scheme combined with all the four transfer scenarios, $M_S$ achieves an ACC of 86.52$\pm$5.61\% for single-channel sleep staging.

\section{Conclusion}
This paper proposed a Multi-Channel Multi-Domain knowledge distillation algorithm for single-channel sleep staging. The proposed algorithm combines knowledge transfer in four different scenarios: Cross-Domain Same-Channel (CDSC), Cross-Domain Cross-Channel (CDCC), Same-Domain Cross-Channel (SDCC) and Same-Domain Same-Channel (SDSC), achieving a state-of-the-art single-channel sleep staging ACC of 86.5\%, with only a 0.6\% deterioration from the state-of-the-art multi-channel model. Experimental results show that the knowledge from multiple domains and multiple channels could be transferred to single-channel EEG sleep staging, bringing an accuracy improvement of 2\%.

\footnotesize

\end{document}